\documentclass[%
reprint,
superscriptaddress,
 amsmath,amssymb,
 aps,
 prl,
floatfix,
]{revtex4-2}
\usepackage{hyperref}

\usepackage{graphicx}
\usepackage{dcolumn}
\usepackage{hyperref}
\usepackage[dvipsnames]{xcolor}
\definecolor{red}{RGB}{250, 0, 0}
\usepackage{bm}
\usepackage{braket}
\usepackage{leftindex}
\graphicspath{{figures/}}
\begin{document}

\preprint{APS/123-QED}

\title{Hanle effect for lifetime determinations in the soft X-ray regime}




\author{Moto Togawa}\email{togawa@mpi-hd.mpg.de}
\affiliation{Max-Planck-Institut f\"ur Kernphysik, Saupfercheckweg 1, 69117 Heidelberg, Germany}%
\affiliation{European XFEL, Holzkoppel 4, 22869 Schenefeld, Germany}%
\affiliation{Heidelberg Graduate School for Physics, Ruprecht-Karls-Universität Heidelberg, Im Neuenheimer Feld 226, 69120 Heidelberg, Germany}%

\author{Jan Richter}%
\affiliation{Physikalisch–Technische Bundesanstalt, D–38116 Braunschweig, Germany}%
\affiliation{Institut f\"ur Theoretische Physik, Leibniz Universität Hannover, Appelstraße 2, 30167 Hannover, Germany}

\author{Chintan Shah}%
\affiliation{NASA/Goddard Space Flight Center, 8800 Greenbelt Road, Greenbelt, MD 20771, USA}%
\affiliation{Max-Planck-Institut f\"ur Kernphysik, Saupfercheckweg 1, 69117 Heidelberg, Germany}%
\affiliation{Department of Physics and Astronomy, Johns Hopkins University, Baltimore, MD 21218, USA}

\author{Marc Botz}%
\affiliation{Max-Planck-Institut f\"ur Kernphysik, Saupfercheckweg 1, 69117 Heidelberg, Germany}%
\affiliation{Heidelberg Graduate School of Fundamental Physics, Ruprecht-Karls-Universität Heidelberg, Im Neuenheimer Feld 226, 69120 Heidelberg, Germany}%

\author{Joshua Nenninger}%
\affiliation{Max-Planck-Institut f\"ur Kernphysik, Saupfercheckweg 1, 69117 Heidelberg, Germany}%

\author{Jonas Danisch}%
\affiliation{Max-Planck-Institut f\"ur Kernphysik, Saupfercheckweg 1, 69117 Heidelberg, Germany}%

\author{Joschka Goes}%
\affiliation{Max-Planck-Institut f\"ur Kernphysik, Saupfercheckweg 1, 69117 Heidelberg, Germany}%

\author{Steffen K\"uhn}%
\affiliation{Max-Planck-Institut f\"ur Kernphysik, Saupfercheckweg 1, 69117 Heidelberg, Germany}%
\affiliation{Heidelberg Graduate School of Fundamental Physics, Ruprecht-Karls-Universität Heidelberg, Im Neuenheimer Feld 226, 69120 Heidelberg, Germany}%

\author{Pedro Amaro}%
\affiliation{%
 Laboratory of Instrumentation, Biomedical Engineering and Radiation Physics (LIBPhys-UNL), Department of Physics, NOVA School of Science and Technology, NOVA University Lisbon, 2829-516 Caparica, Portugal 
}%

\author{Awad Mohamed}%
\affiliation{ISM-CNR, Istituto di Struttura dei Materiali, LD2 Unit, 34149 Trieste, Italy}%
\affiliation{Physics Division, School of Science and Technology, Università di Camerino, Via Madonna delle Carceri 9, Camerino, MC, Italy}

\author{Yuki Amano}%
\affiliation{Institute of Space and Astronautical Science (ISAS), Japan Aerospace Exploration Agency (JAXA), 3-1-1 Yoshinodai, Chuo-ku, Sagamihara, Kanagawa, 252-5210,
Japan}

\author{Stefano Orlando}
\affiliation{ISM-CNR, Istituto di Struttura dei Materiali, LD2 Unit, 34149 Trieste, Italy}

\author{Roberta Totani}%
\affiliation{ISM-CNR, Istituto di Struttura dei Materiali, LD2 Unit, 34149 Trieste, Italy}
\affiliation{Elettra - Sincrotrone Trieste S.C.p.A. Strada Statale 14, 34149 Trieste Italy}

\author{Monica de Simone}%
\affiliation{%
 IOM-CNR, Istituto Officina dei Materiali, 34149 Trieste, Italy
}%
\author{Stephan Fritzsche}
\affiliation{Helmholtz-Institut Jena, Fröbelstieg 3, D-07763 Jena, Germany}
\affiliation{GSI Helmholtzzentrum f\"ur Schwerionenforschung GmbH, Planckstrasse 1, D-64291 Darmstadt, Germany}
\affiliation{Theoretisch-Physikalisches Institut, Friedrich-Schiller-Universität Jena, D-07763 Jena, Germany}

\author{Thomas Pfeifer}
\affiliation{Max-Planck-Institut f\"ur Kernphysik, Saupfercheckweg 1, 69117 Heidelberg, Germany}%

\author{Marcello Coreno}%
\affiliation{ISM-CNR, Istituto di Struttura dei Materiali, LD2 Unit, 34149 Trieste, Italy}

\author{Andrey Surzhykov}%
\affiliation{Physikalisch–Technische Bundesanstalt, D–38116 Braunschweig, Germany}%
\affiliation{Institut f\"ur Mathematische Physik, Technische Universität Braunschweig, Mendelssohnstrasse 3, D-38106 Braunschweig, Germany}

\author{Jos\'e R.~{Crespo L\'opez-Urrutia}}
\affiliation{Max-Planck-Institut f\"ur Kernphysik, Saupfercheckweg 1, 69117 Heidelberg, Germany}%

\date{\today}

\begin{abstract}
By exciting a series of $1\mathrm{s}^{2}\, ^{1}\mathrm{S}_{0} \to 1\mathrm{s}n\mathrm{p}\, ^{1}\mathrm{P}_{1}$ transitions in helium-like nitrogen ions with linearly polarized monochromatic soft X-rays at the Elettra facility, we found a change in the angular distribution of the fluorescence sensitive to the principal quantum number $n$. In particular it is observed that the ratio of emission in directions parallel and perpendicular to the polarization of incident radiation increases with higher $n$. We find this $n$-dependence to be a manifestation of the Hanle effect, which served as a practical tool for lifetime determinations of optical transitions since its discovery in 1924. In contrast to traditional Hanle effect experiments, in which one varies the magnetic field and considers a particular excited state, we demonstrate a 'soft X-ray Hanle effect' which arises in a static magnetic field but for a series of excited states. By comparing experimental data with theoretical predictions, we were able to determine lifetimes ranging from hundreds of femtoseconds to tens of picoseconds of the $1\mathrm{s}n\mathrm{p}\, ^{1}\mathrm{P}_{1}$ levels, which find excellent agreement with atomic-structure calculations. We argue that dedicated soft X-ray measurements could yield lifetime data that is beyond current experimental reach and cannot yet be predicted with sufficient accuracy.
\end{abstract}

                 
\maketitle
Diagnostics of high-temperature plasmas -- whether generated in fusion devices, with high-power lasers from the infrared to the X-ray range, or found in astrophysical observations -- relies on accurate experimental and theoretical data of oscillator strengths for X-ray transitions in highly charged ions (HCI). Generally, few-electron HCI are present in such plasmas. While, in principle, their simpler electronic structure should facilitate calculations and thus diagnostics, there are still large gaps in our experimental knowledge of those systems.
Accurate line positions \cite{leutenegger2020}, oscillator strengths and lifetimes \cite{kuehn2022} were measured for a variety of HCI by exciting them with monochromatized X-rays within electron beam ion traps. 
 Synchrotron radiation facilities provide polarized X-rays to resonantly excite intershell-transitions in HCI, leading to fluorescence emission distributions that depend both on the incident polarisation and the angular momenta of the upper and lower levels. 
 For example, for the simplest case of a $J_i=0 \to J_\nu =1 \to J_f =0$ electric dipole transition, the angular distribution of photons, scattered normal to the direction of the incident linearly polarized beam, exhibits the well known pattern of classical dipole radiation~\cite{balashov2000}:
 \begin{equation}
 W\left(\phi\right)=W_0 \sin^2\phi. 
 \label{Eq.: 1}
 \end{equation} 
 This simple approach suggests, in particular,  no photon emission parallel to the incident polarization vector, i.e. at $\phi=0$.
However, even a weak external magnetic field can affect this distribution, as discovered by Wilhelm Hanle in 1924 \cite{hanle1924}, who showed how it results from the interference of partially overlapping magnetic sublevels.
From then on, the Hanle effect became a key tool for optical lifetime determinations, and is still in use. 
For HCI and also beyond the optical range, however, other experimental methods were necessary (for reviews, see, e.~g., \cite{Traebert2008} on trapped ions, \cite{Traebert1988,Traebert2014} on beam-foil methods, or \cite{Traebert2022,Traebert2022b} for historical overviews). A plethora of theoretical studies has covered lifetimes of allowed and forbidden transitions in many different isoelectronic sequences. 
Lifetimes of long-lived metastable states (e.~g., 1s2s $^{3}$$\mathrm{S}_{1}$) of He-like ions have been repeatedly measured along the isoelectronic sequence with accuracy in some cases better than 1\% \cite{Schmidt_1994,Stefanelli1995,BeiersdorferMagneticTrapping1996,JRCLU_1998,Traebert1999,CrespoHe-likeS2006}. However, fast transitions in heavier elements, e.~g., such as those arising from the $^{1}\mathrm{P}_{1}$ levels, have only been accessible either by beam-foil excitation \cite{Traebert1988} or high-resolution X-ray spectroscopy \cite{Beiersdorfer1996Lifetime-Limited}.
All above techniques are successful in measuring lifetimes within a certain energy and time range. However, no experiment so far has been able to test (soft) X-ray transitions associated with lifetimes in the range from hundreds of femtoseconds to a few picoseconds. To fill this experimental gap, we performed a high-energy analogue of the original Hanle experiment at the soft X-ray beamline, (GasPhase) of the Elettra synchrotron radiation facilty in Trieste, Italy \cite{prince1998, blyth1999}. 
\begin{figure}
    \centering
    \includegraphics[width=0.5\textwidth]{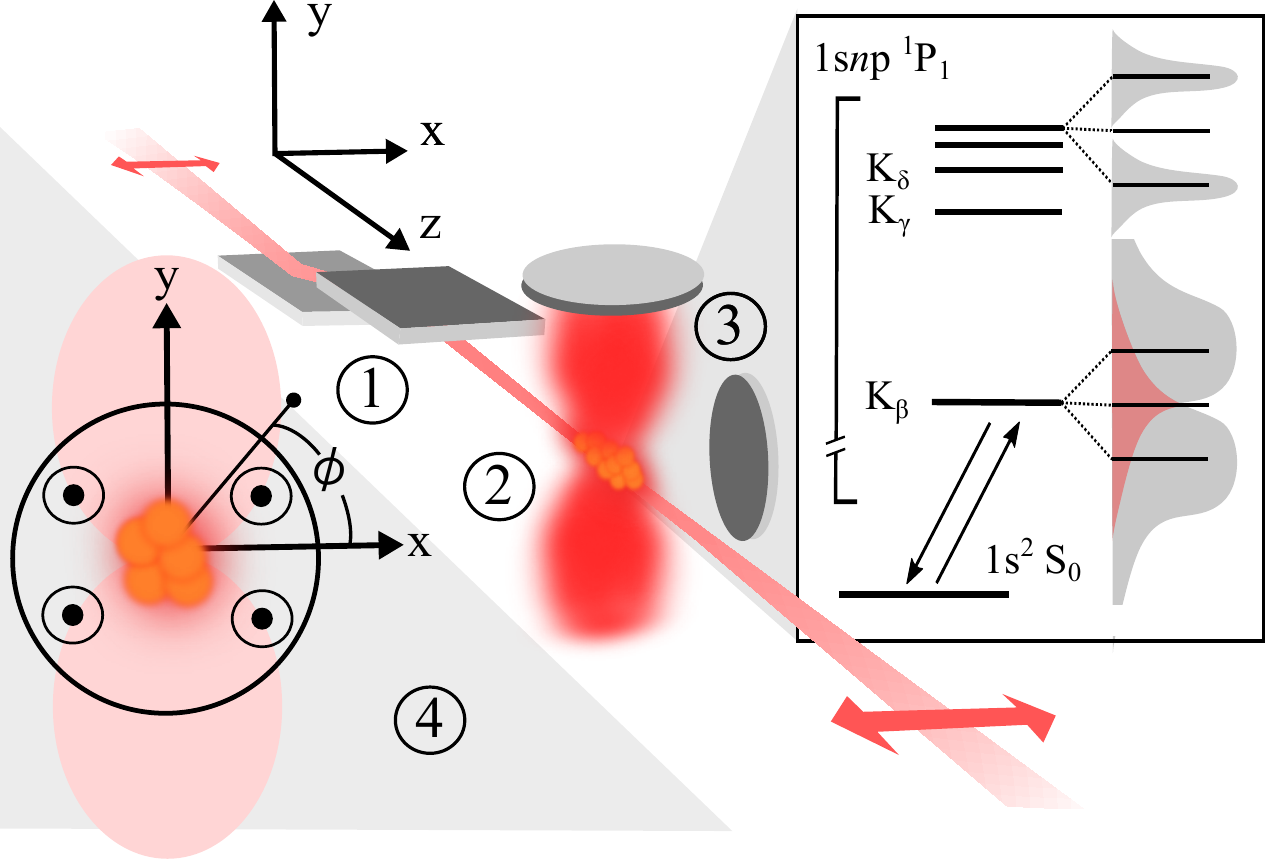}
    \vspace{-8mm}
    \caption{Experimental set up. Linearly polarized undulator radiation is monochromatized (1) and focused on HCIs (2) which are produced and trapped by means of an electron beam (not shown) tightly focused by a magnetic field B. X-rays following resonant excitation are scattered by the ions and recorded by two SDDs (3), mounted one perpendicular and the other parallel to the polarization plane. Their angular distribution (4) is given by Eq.~\ref{Eq.: 1}. Inset left: Grotian diagram of the electronic levels investigated in this work; right: Corresponding magnetic sublevels m$_{j}$. At low \textit{n}, the natural width of several of these states can overlap, allowing for their coherent excitation. At higher \textit{n}, the linewidth decreases, leading to statistical excitation process, i.~e., depolarization.  }
    \label{fig:fig1}
\end{figure}
A portable electron beam ion trap (EBIT), PolarX-EBIT \cite{micke2018}, was installed at the Gasphase beamline. Its off-axis electron gun allows the photon beam to enter and leave the apparatus unimpeded along its longitudinal axis. For the production of the target ions, a tenuous  molecular nitrogen beam is injected into the interaction zone of the EBIT, where highly charged nitrogen ions are produced by means of successive electron impact ionization.
By tuning the electron beam energy to approximately 200 eV, a charge-state distribution dominated by helium-like nitrogen ions is produced and trapped. This electron energy is lower than the excitation threshold for the K-shell, reducing the background caused by electron impact excitation on our signal.
The ions are illuminated by the focused, linearly polarized, monochromatic soft X-ray beam from the GasPhase beamline.
X-rays emitted from the ions are then registered by two silicon-drift detectors (SDD) which are mounted side-on and parallel and perpendicular to the polarization vector of the incident photon beam (see Fig.\ref{fig:fig1}). The fluorescence yield recorded by these two SDDs gives insight into the angular distribution of the emitted fluorescence photons. Their energy resolution of approximately 60\,eV FWHM at the oxygen K-edge additionally allows the fluorescence photons to be distinguished from the background. 
As depicted in Fig. \ref{fig2}, fluorescence is recorded in discrete steps of the incident photon energy. The two detectors record simultaneously the fluorescence yield parallel ($Y_{||}$) and perpendicular ($Y_{\perp}$) to the polarization plane. After integration, we determine the ratio $Y_{||} / Y_{\perp}$. This is repeated for the $1\mathrm{s}^{2} - 1\mathrm{s}n\mathrm{p} ^{1}\mathrm{P}_{1}$ series with \textit{n} up to 7. According to Eq.~\ref{Eq.: 1}, the fluorescence yield $Y_{||}$ should always be equal to zero; finite values are a measure of depolarization. For transitions at higher \textit{n}, we set the monochromator to the corresponding resonances after finding their positions with an initial scan. We then acquire signal for longer times and subtract the EBIT background obtained by closing the photon shutter of the beamline (see Supplemental Material \cite{Supplemental}).
\begin{figure}[t]
    \centering
    \includegraphics[width=0.5\textwidth]{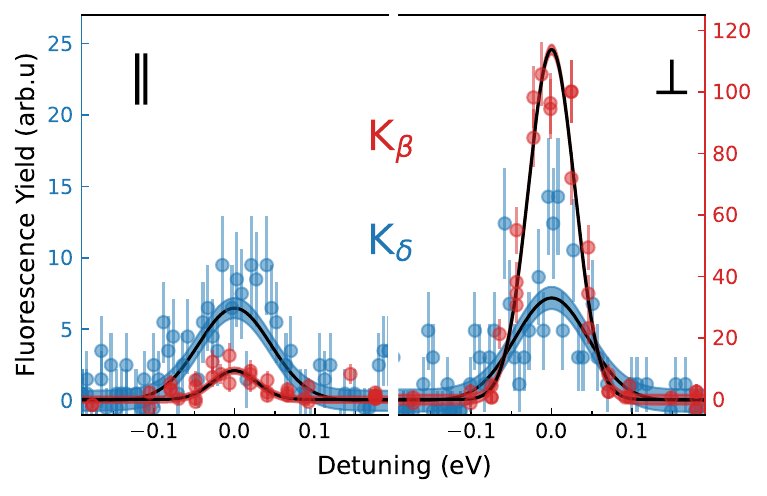}
    \vspace{-9mm}
    \caption{Exemplary measurement of two selected 1$\text{s}$$-n\text{p}$ lines. Left: Fluorescence yield recorded within the polarization plane. Right: Fluorescence yield perpendicular to the polarization plane. The ratio for K$_{\beta}$ is still dominated by the perpendicular contribution in accordance to Eq.~\ref{Eq.: 1}. K$_{\delta}$ shows close to equal emission in both directions.   }
    
    \label{fig2}
\end{figure}

\begin{figure}
    \centering
    \includegraphics[width=0.5\textwidth]{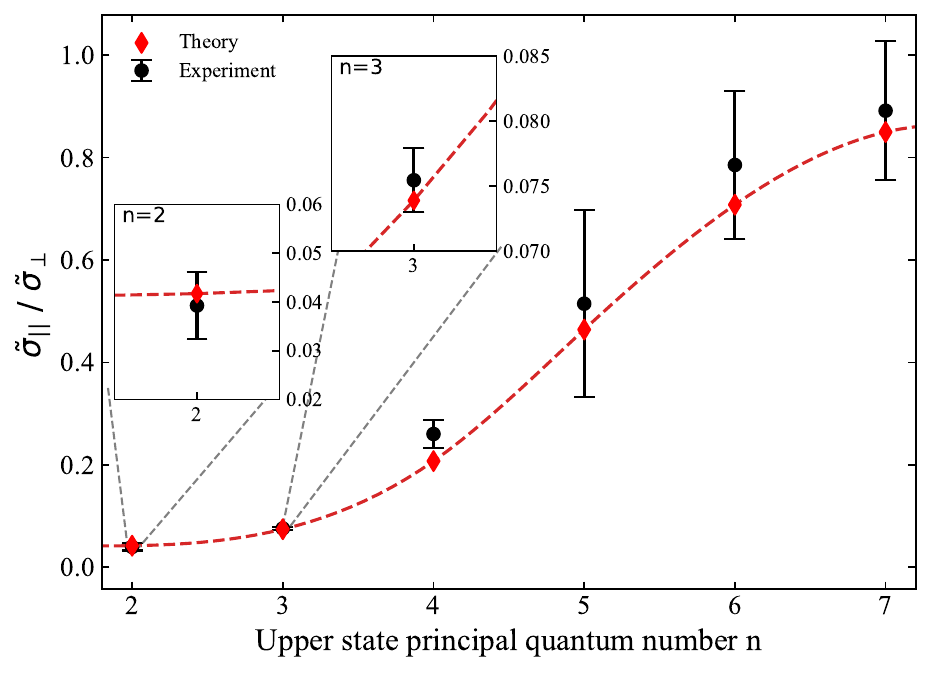}
    \caption{Ratio of cross sections for the resonant scattering of light parallel and perpendicular to the polarization vector of the incident light against the principal quantum number $n$ of the p$_{3/2}$ electron in the $^{1}\mathrm{P}_{1}$ state. Red diamonds represent our theoretical predictions, connected by a dotted red curve to guide the eye. Our calculations of a ratio close to zero for $n = 2$ and increasing for $n > 2$ agree well with the present experimental data. This data is obtained as the ratio of the detected intensities $Y_\parallel/Y_\perp$ which is equal to $\sigma_\parallel/\sigma_\perp$.}
    \label{fig: results experiment and theory}
\end{figure}

By plotting the yield ratios against the principal quantum number $n$ of the upper level, we observe in qualitative agreement with the field-free approximation of Eq.~\ref{Eq.: 1} a ratio close to zero for $n=2$ (K$\alpha$). 
 However, for $n>2$ its value grows rapidly and approaches already at $n = 7$ a value of $\approx 1$, representing full depolarization (Fig.~\ref{fig: results experiment and theory}).
In order to understand these results, we theoretically analyze the resonant elastic scattering of linearly polarized light by evaluating the corresponding second-order matrix element:


\begin{align}
\label{Eq.: Scattering matrix element}
    \mathcal{M}_{M_f,M_i}=\alpha &\sum_{\gamma_\nu J_\nu M_\nu}\left[\frac{\braket{f|\hat{\mathcal{R}}^\dagger\left(\bm{k}_f,\bm{\epsilon}_f\right)|\nu}\braket{\nu|\hat{\mathcal{R}}\left(\bm{k}_i,\bm{\epsilon}_i\right)|i}}{E_i-E_\nu+\omega_i}\right.\notag\\
    &\left.+\frac{\braket{f|\hat{\mathcal{R}}\left(\bm{k}_i,\bm{\epsilon}_i\right)|\nu}\braket{\nu|\hat{\mathcal{R}}^\dagger\left(\bm{k}_f,\bm{\epsilon}_f\right)|i}}{E_i-E_\nu-\omega_i}\right],
\end{align}
with $\alpha$ being the fine structure constant and $\ket{i}=\ket{\gamma_i J_i M_i}$, $\ket{\nu}=\ket{\gamma_\nu J_\nu M_\nu}$, and $\ket{f}=\ket{\gamma_f J_f M_f}$ the short-hand notations for the initial, intermediate, and final electronic states. These states are specified by their total angular momentum $J$, its projection $M$, and $\gamma$, which refers to all additional quantum numbers needed for a unique characterization of the states. For elastic photon scattering, as considered in the present study, the electronic configurations and total angular momentum of initial and final states are the same, $\gamma_i=\gamma_f$ and $J_i=J_f$. The operators $\hat{\mathcal{R}}\left(\bm{k}_i,\bm{\epsilon}_i\right)$ and $\hat{\mathcal{R}}^\dagger\left(\bm{k}_f,\bm{\epsilon}_f\right)$ describe the absorption and emission of photons with wavevectors $\bm{k}_i$ and $\bm{k}_f$ and polarization vectors $\bm{\epsilon}_i$ and $\bm{\epsilon}_f$, respectively~\cite{Serbo22, Volotka22}.

The evaluation of Eq.~\eqref{Eq.: Scattering matrix element} requires a summation over the complete atomic spectrum $\ket{\gamma_\nu J_\nu M_\nu}$. This sum can be truncated to a single term when the energy of the incident photon is close to a transition between initial and one of the intermediate states $\omega_i\approx E_\nu - E_i$. In this resonant case, the scattering amplitude can be simplified to:
\begin{equation}
\label{Eq.: Scattering matrix element resonant}
    \mathcal{M}^{res}_{M_f,M_i} \approx \alpha \sum_{M_\nu}\frac{\braket{f|\hat{\mathcal{R}}^\dagger\left(\bm{k}_f,\bm{\epsilon}_f\right)|\nu}\braket{\nu|\hat{\mathcal{R}}\left(\bm{k}_i,\bm{\epsilon}_i\right)|i}}{E_i-E_\nu+\omega_i+i\Gamma_\nu/2}.
\end{equation}
Here, the natural width $\Gamma_\nu$ of the intermediate state was phenomenologically introduced to the denominator to avoid a divergency for the case of a zero energy detuning, i.e. when $\omega_i=E_\nu-E_i$~\cite{Samoilenko2020, Serbo22, Volotka22}.
Using the matrix element~\eqref{Eq.: Scattering matrix element resonant} one can obtain the differential cross section 
\begin{equation}
\label{Eq: differential cross section}
     \frac{\mathrm{d}\sigma}{\mathrm{d}\Omega}\left(\theta_f,\phi_f\right) = \frac{1}{2 J _i +1} \sum_{M_i ,M_f, \bm{\epsilon}_f}\left|\mathcal{M}^{res}_{M_f,M_i}\right|^2,
 \end{equation}
for the resonant scattering of a photon under the angles $\left(\theta_f,\phi_f\right)$ defined with respect to the direction $\bm{k}_i$ and polarization $\bm{\epsilon}_i$ of the incident radiation. Here, we assumed an unpolarized initial state of the ion and that both the population of the magnetic sublevels $\ket{\gamma_f J_f M_f}$ as well as the polarization of the scattered light remain unobserved.

In what follows, we apply Eqs. \eqref{Eq.: Scattering matrix element resonant}-\eqref{Eq: differential cross section} to analyze the angular distribution of the $1\mathrm{s}^2 \leftindex^1 {\mathrm{S}}_0 \to 1\mathrm{s} n\mathrm{p}\leftindex^1{\mathrm{P}}_1 \to 1\mathrm{s}^2 \leftindex^1 {\mathrm{S}}_0$ scattering of initially linearly polarized photons. By considering the experimental setup from Fig.~\ref{fig:fig1}, with scattered photons detected normal to the propagation direction of the incident radiation $\left(\theta_f=\pi/2\right)$ and either within $\left(\phi_f=0\right)$ or perpendicular $\left(\phi_f=\pi/2\right)$ to its polarization axis, we find:
\begin{subequations}
\begin{align}
    \sigma_\parallel &\equiv  \frac{\mathrm{d}\sigma}{\mathrm{d}\Omega}\left(\theta_f=\frac{\pi}{2},\phi_f=0\right) =0, \label{eq: cross section parallel B=0}\\
    \sigma_\perp &\equiv  \frac{\mathrm{d}\sigma}{\mathrm{d}\Omega}\left(\theta_f=\frac{\pi}{2},\phi_f=\frac{\pi}{2}\right) =\frac{4R^2}{{\Gamma_\nu ^2}+4\Delta\omega ^2} \label{eq: cross section perpendicular B=0}.
\end{align}
\end{subequations}
Here, $R=2\pi \alpha \left|\braket{1\mathrm{s} n\mathrm{p}\leftindex^1{\mathrm{P}}_1||\hat{a}_{E1}||1\mathrm{s}^2 \leftindex^1 {\mathrm{S}}_0}\right|^2$ is the square of the reduced matrix element of the electric dipole transition $1\mathrm{s}^2 \leftindex^1 {\mathrm{S}}_0 \to 1\mathrm{s} n\mathrm{p}\leftindex^1{\mathrm{P}}_1$ and $\Delta \omega = E_\nu-E_i-\omega$ is the detuning between the incident photon energy and the transition energy.
As seen from Eqs.~\eqref{eq: cross section parallel B=0}-\eqref{eq: cross section perpendicular B=0}, the standard second-order perturbation approach negates photon scattering parallel to the polarization vector of the incident light, $\sigma_\parallel=0$, and hence, predicts a vanishing ratio $\sigma_\parallel/\sigma_\perp$, in contradiction with our experimental findings.\\
So far, our theoretical analysis was restricted to the case of resonant scattering in the absence of external electromagnetic fields. However, in our experiment the ions are perturbed by the magnetic field of the EBIT. This field is aligned with the propagation direction of the incident light, chosen as the z-axis, and leads to a Zeeman splitting of the electronic levels. In a similar manner as in Ref.~\cite{Stenflo1998}, the Zeeman effect can be incorporated into the perturbative approach, leading to a modified scattering amplitude:
\begin{equation}
\label{Eq: scattering matrix element B field}
    \mathcal{M}^{res}_{M_f,M_i} \approx \alpha \sum_{M_\nu}\frac{\braket{f|\hat{\mathcal{R}}^\dagger\left(\bm{k}_f,\bm{\epsilon}_f\right)|\nu}\braket{\nu|\hat{\mathcal{R}}\left(\bm{k}_i,\bm{\epsilon}_i\right)|i}}{E_i-\left(E_\nu+M_\nu \Delta E_Z\right)+\omega_i +i\Gamma_\nu/2},
\end{equation}
with $\Delta E_Z = g_j \mu_B B$ and the unperturbed energy of the intermediate state $E_\nu$. One may note that Eq. \eqref{Eq: scattering matrix element B field} is obtained under the assumption that the initial (and, hence, final) state does not exhibit any Zeeman splitting since $J_i=J_f=0$.

By applying Eqs.~\eqref{Eq: differential cross section} and \eqref{Eq: scattering matrix element B field} for the $1\mathrm{s}^2 \leftindex^1 {\mathrm{S}}_0 \to 1\mathrm{s} n\mathrm{p}\leftindex^1{\mathrm{P}}_1 \to 1\mathrm{s}^2 \leftindex^1 {\mathrm{S}}_0$ scattering, we derive the cross sections
\begin{subequations}
\label{eq: cross sections B field}
\begin{align}
\sigma_\parallel & =16 \kappa \Delta E_Z^2,\\
\sigma_\perp & =4\kappa  \left(\Gamma_\nu^2+4\Delta\omega^2\right),
\end{align}
\end{subequations}
for photons scattered parallel and perpendicular to the polarization $\bm{\epsilon}_i$ of the incident light, with the parameter $\kappa~=~R^2/{\left[{\Gamma_\nu^2}+4\left(\Delta\omega-E_z\right)^2\right]\left[{\Gamma_\nu^2}+4\left(\Delta\omega+E_z\right)^2\right]}$.
With the help of Eq.~\eqref{eq: cross sections B field}, we obtain the ratio: 
\begin{equation}
\label{Eq.: cross section ratio}
     \frac{\sigma_\parallel}{\sigma_\perp}=\frac{4 \Delta E_Z^2}{\Gamma_\nu^2+4\Delta\omega^2},
\end{equation}
which can deviate from zero for an ion in an external magnetic field. The effect depends on experimental parameters and ion properties, and will be discussed later. We just mention here that Eq.~\eqref{Eq.: cross section ratio} corresponds to the low-intensity limit of the non-perturbative treatment of the Hanle effect (see Ref.~\cite{Avan1975} for further details).

Expression \eqref{Eq.: cross section ratio} was derived for ideally monochromatic and completely polarized incident radiation as well as for point-like detectors, and has to be modified for realistic conditions. Since the width of the incident radiation $\Delta\omega\approx 0.1$ eV is larger than the Zeeman splitting of the intermediate sublevels, we use an energy-averaged cross section:
\begin{equation}
    \tilde\sigma_{\perp,\parallel} = \int \sigma_{\perp,\parallel}\left(\omega\right) \mathcal{G}\left(\omega\right) \mathrm{d}\omega.
    \label{Eq.: Averaged corss section}
\end{equation}
Here, $\mathcal{G}\left(\omega\right)$ is a Gaussian distribution, and the incident synchrotron radiation is assumed to be incoherent. We also take into account the detector size by integrating the differential cross section~\eqref{Eq: differential cross section} over a finite solid angle. Finally, the effect of incomplete polarization of the incident radiation is considered within the density-matrix approach. Details are given in the Supplemental Material \cite{Supplemental}.
\begin{table}[b]
    \centering
    \begin{tabular}{|l|l|l|l|}
    \hline
           & \multicolumn{2}{c}{Theory} & Experiment \\
         $n$ \hspace*{0.1cm}& Energy (eV) & Lifetime (ps) & Lifetime (ps) \\
         \hline
         2 &$430.73$                     &$0.5539$                    &$(0.58 \pm 0.58)$\\
           &$430.71$ \cite{johnson_2002} &$0.5537$ \cite{johnson_2002} & \\
           &$430.55$ \cite{Cann_1992}    &$0.5531$ \cite{Cann_1992} & \\
          &&&\\
         3 &$497.95$                     &$1.844$                      &$(1.98 \pm 0.17)$ \\
           &$497.93$ \cite{johnson_2002} &$1.846$ \cite{johnson_2002} & \\
           &$497.75$ \cite{Cann_1992}    &$1.844$ \cite{Cann_1992} & \\
          &&&\\
         4 &$521.60	$                   &$4.346$                      &$(5.3 \pm 0.7)$\\
           &$521.58$ \cite{johnson_2002} &$4.346$ \cite{johnson_2002} & \\
           &$521.39$ \cite{Cann_1992}   & $4.337$ \cite{Cann_1992} & \\
          &&&\\
         5 &$532.56$                     &$8.445$                      &$(9.78 \pm 4.46)$ \\
           &$532.56$ \cite{johnson_2002} &$8.449$ \cite{johnson_2002} & \\
           &$532.35$ \cite{Cann_1992}    &$8.423$ \cite{Cann_1992} & \\
          &&&\\
         6 &$538.53$                     &$14.51$                      &$(19.89 \pm 11.85)$\\
          & $538.50$ \cite{johnson_2002} &$14.44$ \cite{johnson_2002} & \\
          &$538.31$ \cite{Cann_1992}     &$14.51$ \cite{Cann_1992} & \\
           &&&\\
         7 & $542.13$                    &$22.79$                      &$(35.29 \pm 44.79)$\\
         \hline
    \end{tabular}
    \caption{The excitation energies, defined with respect to the ground $1\mathrm{s}^2\leftindex^1{\mathrm{S}}_0$ state, and lifetimes of the $1\mathrm{s} n\mathrm{p}\leftindex^1{\mathrm{P}}_1$ levels of a $\mathrm{N}^{5+}$ ion. Theoretical predictions obtained using the AMBiT code are compared with previous calculations from Refs.~\cite{johnson_2002,Cann_1992}, and with present experimental findings for the lifetimes.
    }
    \label{tab: theory results}
\end{table}

\begin{figure}
    \centering
    \includegraphics[width=0.5\textwidth]{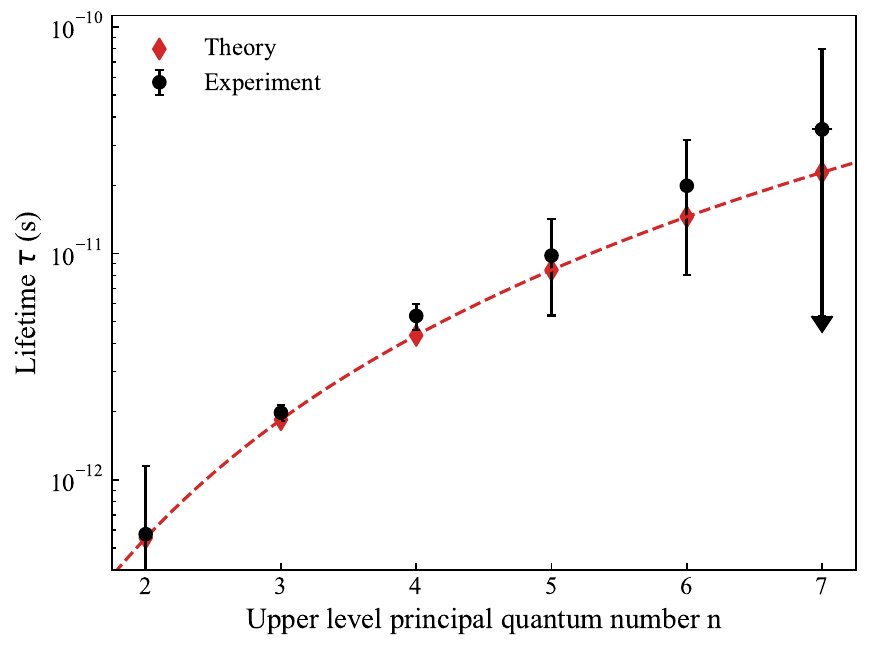}
    \caption{Lifetime $\tau = \hbar/\Gamma_\nu$ of the $1\mathrm{s} n\mathrm{p}\leftindex^1{\mathrm{P}}_1$ states depending on the principal quantum number $n$. Our experimental data (black dots) is compared with theoretical predictions obtained using the AMBiT code (red diamonds), connected by a dotted red curve to guide the eye.  
    }
    \label{fig: level width}
\end{figure}


We now apply the perturbative approach discussed above to analyze our experimental findings. We start from Eq.~\eqref{Eq.: cross section ratio} which predicts that the ratio $\sigma_\parallel/\sigma_\perp$ might deviate from zero due to a Zeeman splitting of the intermediate state, and can be used to explain the pronounced $n$-dependence. Since the g-factor, and hence the Zeeman shift, are almost constant for the entire series $1\mathrm{s} n\mathrm{p}\leftindex^1{\mathrm{P}}_1$~\cite{Puchalski2012}, this strong dependence must arise from the lifetime $\tau$ and, hence, the natural width $\Gamma_\nu=\hbar/\tau$ of the excited states. Indeed, as seen from Table \ref{tab: theory results}, $\tau$ is very sensitive to the principal quantum number $n$. For example, the lifetime of the $1\mathrm{s} 7\mathrm{p}\leftindex^1{\mathrm{P}}_1$ state is increased by a factor of 40 compared to $\tau\left(1\mathrm{s} 2\mathrm{p}\leftindex^1{\mathrm{P}}_1\right)$.
These results were obtained with the configuration-interaction method implemented in the AMBiT code~\cite{KAHL2019}, and agree well with previous calculations from Refs.~\cite{johnson_2002,Cann_1992}.\\
The remarkable prolongation of the lifetime as a function of $n$, and the concomitant reduction of the natural width $\Gamma_\nu=\hbar/\tau$ leads to the growth of the ratio $\sigma_\parallel/\sigma_\perp$, as we observe in the experiment.
For a quantitative comparison with experimental data we make use of the energy-averaged cross section~\eqref{Eq.: Averaged corss section} and take into account corrections due to the finite size of the photon detectors. Moreover, we assume complete linear polarization of the incident synchrotron radiation (see Refs.~\cite{Rouvellou2003,karvonen1999,micke2018}) and take into account the EBIT magnetic field of $B = 0.85$~T. For these parameters, we calculate the modified cross section ratio $\tilde \sigma_\parallel/\tilde\sigma_\perp$ shown by the red diamonds in Fig.~\ref{fig: results experiment and theory}. The  good agreement of theory and experiment supports our explanation of the effect based on the Zeeman splitting of ionic levels. The clear $n$-dependence of the ratio $\sigma_\parallel/\sigma_\perp$ manifests the well-known Hanle effect, now found in the soft X-ray domain.\\
One may note, that our setup for the observation of the Hanle effect differs from the traditional one. Indeed for the latter, one varies the external magnetic field to change the overlap of magnetic sublevels of a specific intermediate state $\ket{\gamma_\nu J_\nu}$. This, in turn, alters the interference contributions of the substates $\ket{\gamma_\nu J_\nu M_\nu}$ to the scattering cross section and, hence, leads to a modification of polarization and angular distribution of outgoing photons. In contrast, the external magnetic field remains constant in our experiment. The variation of the overlap of the Zeeman sublevels and, hence, of their inteference contributions is achieved by adressing various $1\mathrm{s} n\mathrm{p}\leftindex^1{\mathrm{P}}_1$ states exhibiting different natural widths $\Gamma_\nu$.
This results in analogous modifications to the angular distribution, effectively demonstrating the Hanle effect through the $n$-dependence in our setup.
With these insights, we can extract information about the lifetimes of the intermediate $1\mathrm{s} n\mathrm{p}\leftindex^1{\mathrm{P}}_1$ states from the measured ratio $\sigma_\parallel/\sigma_\perp$. To achieve this, we fit our theoretical predictions of the modified cross section ratio $\tilde\sigma_\parallel/\tilde\sigma_\perp$ to the experimental data from Fig.~\ref{fig: results experiment and theory}. The only free parameter for the fitting procedure is $\Gamma_\nu$, while all other experimental parameters are taken as in the calculations above. The uncertainty of the derived lifetimes $\tau\left(1\mathrm{s} n\mathrm{p}\leftindex^1{\mathrm{P}}_1\right)$, beeing about few tens of percent, is estimated by propagating those of the measured cross-section ratios and the magnetic field strength (see Supplemental Material \cite{Supplemental}). The derived results, displayed in Fig.~\ref{fig: level width}, show good agreement with the theoretical predictions.

In conclusion, our present method based on the Hanle effect yields lifetimes of excited levels from abundant species of highly charged ions in the soft X-ray regime. This method is feasible for a range of hundreds of femtoseconds to tens of picoseconds for which resolution of linewidths is still beyond reach at advanced light sources. Our present experimental accuracy will improve with 
more statistics in future campaigns. Following this demonstration with theoretically well-understood He-like ions, we will apply our method to the Li-like and other isoelectronic sequences, in which transitions, broadened by more complex fast autoionization channels, still challenge theory. Experimental results for such transitions are urgently needed for an improved scientific harvest of X-ray space observatory data from Chandra, XMM-Newton, and the recently launched XRISM \cite{xrism}. Furthermore, applications for magnetic field and polarization studies in hot plasmas are expected \cite{Fineschi1993,sahal1986}.

\begin{acknowledgments}
Financial support was provided by the Max-Planck-Gesellschaft (MPG), Bun\-des\-mi\-ni\-ste\-ri\-um f{\"u}r Bildung und Forschung (BMBF) through project 05K13SJ2 and by the AHEAD-2020 Project grant agreement 871158 of the European Union’s Horizon 2020 Programme. C.S. acknowledges support from NASA under award number 80GSFC21M0002 and Max-Planck-Gesellschaft (MPG). P.A. acknowledges support from Funda\c{c}\~{a}o para a Ci\^{e}ncia e Tecnologia, Portugal, under grant No.~UID/FIS/04559/2020 (LIBPhys). We acknowledge Elettra Sincrotrone Trieste for providing access to its synchrotron radiation facilities (long term proposals nr.20205206). Fabio Zuccaro is acknowledged for technical support during beamtime preparation and realization. J.R. and A.S. acknowledge funding by the Deutsche Forschungsgemeinschaft (DFG, German Research Foundation) under Germany’s Excellence Strategy – EXC-2123 QuantumFrontiers – 390837967.
\end{acknowledgments}


\begin{thebibliography}{10}

\bibitem{leutenegger2020}
M.~A. Leutenegger, S.~K\"uhn, P.~Micke, R.~Steinbr\"ugge, J.~Stierhof, C.~Shah, N.~Hell, M.~Bissinger, M.~Hirsch, R.~Ballhausen, M.~Lang, C.~Gr\"afe, S.~Wipf, R.~Cumbee, G.~L. Betancourt-Martinez, S.~Park, V.~A. Yerokhin, A.~Surzhykov, W.~C. Stolte, J.~Niskanen, M.~Chung, F.~S. Porter, T.~St\"ohlker, T.~Pfeifer, J.~Wilms, G.~V. Brown, J.~R. Crespo L\'opez-Urrutia, and S.~Bernitt.
\newblock High-precision determination of oxygen ${K}_{\ensuremath{\alpha}}$ transition energy excludes incongruent motion of interstellar oxygen.
\newblock {\em Phys. Rev. Lett.}, 125:243001, Dec 2020.

\bibitem{kuehn2022}
Steffen K\"uhn, Charles Cheung, Natalia~S. Oreshkina, Ren\'e Steinbr\"ugge, Moto Togawa, Sonja Bernitt, Lukas Berger, Jens Buck, Moritz Hoesch, J\"orn Seltmann, Florian Trinter, Christoph~H. Keitel, Mikhail~G. Kozlov, Sergey~G. Porsev, Ming~Feng Gu, F.~Scott Porter, Thomas Pfeifer, Maurice~A. Leutenegger, Zolt\'an Harman, Marianna~S. Safronova, Jos\'e R.~Crespo L\'opez-Urrutia, and Chintan Shah.
\newblock New measurement resolves key astrophysical {Fe XVII} oscillator strength problem.
\newblock {\em Phys. Rev. Lett.}, 129:245001, Dec 2022.

\bibitem{balashov2000}
Vsevolod~V Balashov, Alexei~N Grum-Grzhimailo, and Nikolai~M Kabachnik.
\newblock {\em Polarization and correlation phenomena in atomic collisions: a practical theory course}.
\newblock Springer Science \& Business Media, 2000.

\bibitem{hanle1924}
Wilhelm Hanle.
\newblock {\"U}ber magnetische {B}eeinflussung der {P}olarisation der {R}esonanzfluoreszenz.
\newblock {\em Zeitschrift f{\"u}r Physik}, 30(1):93--105, 1924.

\bibitem{Traebert2008}
E~Träbert.
\newblock Atomic lifetime measurements employing an electron beam ion trap.
\newblock {\em Canadian Journal of Physics}, 86(1):73--97, 2008.

\bibitem{Traebert1988}
E.~Tr\"{a}bert, P.~H. Heckmann, R.~Hutton, and I.~Martinson.
\newblock Intercombination lines in delayed beam-foil spectra.
\newblock {\em J. Opt. Soc. Am. B}, 5(10):2173--2182, Oct 1988.

\bibitem{Traebert2014}
Elmar Tr{\"a}bert.
\newblock Measurement of femtosecond atomic lifetimes using ion traps.
\newblock {\em Applied Physics B}, 114(1):167--172, Jan 2014.

\bibitem{Traebert2022}
Elmar Träbert.
\newblock Atomic lifetime data and databases.
\newblock {\em Atoms}, 10(2), 2022.

\bibitem{Traebert2022b}
Elmar Träbert.
\newblock On atomic lifetimes and environmental density.
\newblock {\em Atoms}, 10(4), 2022.

\bibitem{Schmidt_1994}
H.~T. Schmidt, P.~Forck, M.~Grieser, D.~Habs, J.~Kenntner, G.~Miersch, R.~Repnow, U.~Schramm, T.~Sch\"ussler, D.~Schwalm, and A.~Wolf.
\newblock High-precision measurement of the magnetic-dipole decay rate of metastable heliumlike carbon ions in a storage ring.
\newblock {\em Phys. Rev. Lett.}, 72:1616--1619, Mar 1994.

\bibitem{Stefanelli1995}
G.~S. Stefanelli, P.~Beiersdorfer, V.~Decaux, and K.~Widmann.
\newblock Measurement of the radiative lifetime of the 1s2s${\mathrm{}}^{3}${${\mathrm{S}}_{1}$} level in heliumlike magnesium.
\newblock {\em Phys. Rev. A}, 52:3651--3654, Nov 1995.

\bibitem{BeiersdorferMagneticTrapping1996}
P.~Beiersdorfer, L.~Schweikhard, J.~{Crespo López‐Urrutia}, and K.~Widmann.
\newblock {The magnetic trapping mode of an electron beam ion trap: New opportunities for highly charged ion research}.
\newblock {\em Review of Scientific Instruments}, 67(11):3818--3826, 11 1996.

\bibitem{JRCLU_1998}
J.~R. Crespo L\'opez-Urrutia, P.~Beiersdorfer, D.~W. Savin, and K.~Widmann.
\newblock {Precision measurement of the lifetime of the $1\mathrm{s}2\mathrm{s}{}^{3}{\mathrm{S}}_{1}$ metastable level in heliumlike ${\mathrm{O}}^{6+}$}.
\newblock {\em Phys. Rev. A}, 58:238--241, Jul 1998.

\bibitem{Traebert1999}
E.~Tr\"abert, P.~Beiersdorfer, G.~V. Brown, A.~J. Smith, S.~B. Utter, M.~F. Gu, and D.~W. Savin.
\newblock Improved electron-beam ion-trap lifetime measurement of the {${\mathrm{Ne}}^{8+} 1\mathrm{s}2\mathrm{s}{}^{3}{\mathrm{S}}_{1}$} level.
\newblock {\em Phys. Rev. A}, 60:2034--2038, Sep 1999.

\bibitem{CrespoHe-likeS2006}
J.~R. Crespo L\'opez-Urrutia, P.~Beiersdorfer, and K.~Widmann.
\newblock {Lifetime of the $1\mathrm{s}2\mathrm{s}\phantom{\rule{0.2em}{0ex}}^{3}\mathrm{S}_{1}$ metastable level in $\mathrm{He}$-like ${\mathrm{S}}^{14+}$ measured with an electron beam ion trap}.
\newblock {\em Phys. Rev. A}, 74:012507, Jul 2006.

\bibitem{Beiersdorfer1996Lifetime-Limited}
P.~Beiersdorfer, A.~L. Osterheld, V.~Decaux, and K.~Widmann.
\newblock Observation of lifetime-limited x-ray linewidths in cold highly charged ions.
\newblock {\em Phys. Rev. Lett.}, 77:5353--5356, Dec 1996.

\bibitem{prince1998}
KC~Prince, RR~Blyth, R~Delaunay, M~Zitnik, J~Krempasky, J~Slezak, R~Camilloni, L~Avaldi, M~Coreno, Gly Stefani, et~al.
\newblock The gas-phase photoemission beamline at elettra.
\newblock {\em Journal of synchrotron radiation}, 5(3):565--568, 1998.

\bibitem{blyth1999}
R.R Blyth, R~Delaunay, M~Zitnik, J~Krempasky, R~Krempaska, J~Slezak, K.C Prince, R~Richter, M~Vondracek, R~Camilloni, L~Avaldi, M~Coreno, G~Stefani, C~Furlani, M~{de Simone}, S~Stranges, and M.-Y Adam.
\newblock The high resolution gas phase photoemission beamline, elettra.
\newblock {\em Journal of Electron Spectroscopy and Related Phenomena}, 101-103:959--964, 1999.

\bibitem{micke2018}
P.~Micke, S.~Kühn, L.~Buchauer, J.~R. Harries, T.~M. Bücking, K.~Blaum, A.~Cieluch, A.~Egl, D.~Hollain, S.~Kraemer, T.~Pfeifer, P.~O. Schmidt, R.~X. Schüssler, Ch. Schweiger, T.~Stöhlker, S.~Sturm, R.~N. Wolf, S.~Bernitt, and J.~R. Crespo López-Urrutia.
\newblock {The Heidelberg compact electron beam ion traps}.
\newblock {\em Review of Scientific Instruments}, 89(6):063109, 06 2018.

\bibitem{Supplemental}
\url{URL_will_be_inserted_by_publisher}, 2024.

\bibitem{Serbo22}
V.~G. Serbo, A.~Surzhykov, and A.~Volotka.
\newblock Resonant scattering of plane-wave and twisted photons at the gamma factory.
\newblock {\em Ann. Phys.(Berlin)}, 534(2100199), 2022.

\bibitem{Volotka22}
A.~Volotka, D.~Samoilenko, S.~Fritzsche, V.~G. Serbo, and A.~Surzhykov.
\newblock Polarization of photons scattered by ultra-relativistic ion beams.
\newblock {\em Ann. Phys.(Berlin)}, 534(2100252), 2022.

\bibitem{Samoilenko2020}
D.~Samoilenko, A.~V. Volotka, and S.~Fritzsche.
\newblock Elastic photon scattering on hydrogenic atoms near resonances.
\newblock {\em Atoms}, 8(2), 2020.

\bibitem{Stenflo1998}
J.~O. Stenflo.
\newblock Hanle-zeeman scattering matrix.
\newblock {\em Astronomy and Astrophysics}, 338:301--310, 1998.

\bibitem{Avan1975}
P.~Avan and C.~Cohen-Tannoudji.
\newblock Hanle effect for monochromatic excitation. non perturbative calculation for a {$J = 0$ to $J = 1$} transition.
\newblock {\em Journal de Physique Lettres}, 36(4):85—88, 1975.

\bibitem{johnson_2002}
WR~Johnson, IM~Savukov, UI~Safronova, and A~Dalgarno.
\newblock E1 transitions between states with $n$= 1-6 in helium-like carbon, nitrogen, oxygen, neon, silicon, and argon.
\newblock {\em The Astrophysical Journal Supplement Series}, 141(2):543, 2002.

\bibitem{Cann_1992}
Natalie~Mary Cann and Ajit~J. Thakkar.
\newblock Oscillator strengths for {S-P and P-D} transitions in heliumlike ions.
\newblock {\em Phys. Rev. A}, 46:5397--5405, Nov 1992.

\bibitem{Puchalski2012}
M.~Puchalski and U.~D. Jentschura.
\newblock Quantum electrodynamic corrections to the $g$ factor of helium {$\mathrm{P}$} states.
\newblock {\em Phys. Rev. A}, 86:022508, Aug 2012.

\bibitem{KAHL2019}
E.V. Kahl and J.C. Berengut.
\newblock ambit: A programme for high-precision relativistic atomic structure calculations.
\newblock {\em Computer Physics Communications}, 238:232--243, 2019.

\bibitem{Rouvellou2003}
B.~Rouvellou, S.~Rioual, L.~Avaldi, R.~Camilloni, G.~Stefani, and G.~Turri.
\newblock Angle-dependent postcollisional interaction and interference effects in resonant double photoionization of neon.
\newblock {\em Phys. Rev. A}, 67:012706, Jan 2003.

\bibitem{karvonen1999}
J.~Karvonen, A.~Kivim\"aki, H.~Aksela, S.~Aksela, R.~Camilloni, L.~Avaldi, M.~Coreno, M.~de~Simone, and K.~C. Prince.
\newblock Angular distribution in xenon ${M}_{4,5}{N}_{4,5}{N}_{4,5}$ auger decay.
\newblock {\em Phys. Rev. A}, 59:315--319, Jan 1999.

\bibitem{xrism}
M.~{Tashiro}, H.~{Maejima}, K.~{Toda}, R.~{Kelley}, L.~{Reichenthal}, J.~{Lobell}, R.~{Petre}, M.~{Guainazzi}, E.~{Costantini}, M.~{Edison}, et~al.
\newblock {Concept of the X-ray Astronomy Recovery Mission}.
\newblock {\em Proc. SPIE}, 10699:1069922, 2018.

\bibitem{Fineschi1993}
Silvano Fineschi, Richard~B. Hoover, Muamer Zukic, Jongmin Kim, Arthur B. C.~Walker II, and Phillip~C. Baker.
\newblock {Polarimetry of HI Lyman-alpha for coronal magnetic field diagnostics}.
\newblock In Richard~B. Hoover and Arthur B. C.~Walker II, editors, {\em Multilayer and Grazing Incidence X-Ray/EUV Optics for Astronomy and Projection Lithography}, volume 1742, pages 423 -- 438. International Society for Optics and Photonics, SPIE, 1993.

\bibitem{sahal1986}
S~Sahal-Br{\'e}chot, M~Malinovsky, and V~Bommier.
\newblock The polarization of the o vi 1032 a line as a probe for measuring the coronal vector magnetic field via the hanle effect.
\newblock {\em Astronomy and Astrophysics (ISSN 0004-6361), vol. 168, no. 1-2, Nov. 1986, p. 284-300.}, 168:284--300, 1986.

\end{thebibliography}

\end{document}